\documentclass[doublespace,sagev]{sagej}
\usepackage{fullpage,amsmath,amssymb,dsfont,color,multirow}
\usepackage{setspace}
\usepackage{graphicx}

\usepackage{graphicx, psfrag, amsfonts, bm}
\usepackage{graphics}
\usepackage{amsfonts}
\usepackage{comment}
\usepackage{lipsum}
\usepackage{lscape}
\usepackage[normalem]{ulem}
\usepackage{arydshln}
\usepackage{url}

\graphicspath{ {./figures/} }

\newcommand{\AK}{\textcolor{black}}
\newcommand{\pari}{\hspace{\parindent}}

\newcommand{\bc}{\begin{center}}
\newcommand{\ec}{\end{center}}
\newcommand{\beq}{\begin{equation}}
\newcommand{\eeq}{\end{equation}}
\newcommand{\bea}{\begin{eqnarray}}
\newcommand{\eea}{\end{eqnarray}}
\newcommand{\beas}{\begin{eqnarray*}}
\newcommand{\eeas}{\end{eqnarray*}}
\newcommand{\bi}{\begin{itemize}}
\newcommand{\ei}{\end{itemize}}

\def\beq{\begin{equation}}
\def\eeq{\end{equation}}

\begin{document}

\runninghead{Kaizer et al.}

\title{Statistical design considerations for trials that study multiple indications}

\author{Alexander M. Kaizer\affilnum{1}, Joseph S. Koopmeiners\affilnum{2}, Nan Chen\affilnum{3}, Brian P. Hobbs\affilnum{4} }

\affiliation{\affilnum{1}Department of Biostatistics and Informatics,  University of Colorado-Anschutz Medical Campus, Aurora, CO\\
\affilnum{2}Division of Biostatistics, University of Minnesota, Minneapolis, Minnesota\\
\affilnum{3}Department of Biostatistics,  University of Texas M.D. Anderson Cancer Center, Houston, TX\\
\affilnum{4}Quantitative Health Sciences; Taussig Cancer Institute, Cleveland Clinic, Cleveland, OH
}

\corrauth{Alex Kaizer, 13001 E. 17th Place, Mail Stop B119, Aurora, CO 80045}
\email{alex.kaizer@cuanschutz.edu}

\begin{abstract}
Breakthroughs in cancer biology have defined new research programs emphasizing the development of therapies that target specific pathways in tumor cells. Innovations in clinical trial design have followed with master protocols defined by inclusive eligibility criteria and evaluations of multiple therapies and/or histologies. Consequently, characterization of subpopulation heterogeneity has become central to the formulation and selection of a study design. \AK{However, this transition to master protocols has led to challenges in identifying the optimal trial design and proper calibration of hyperparameters. We often evaluate a range of null and alternative scenarios, however there has been little guidance on how to synthesize the potentially disparate recommendations for what may be optimal. This may lead to the selection of suboptimal designs and statistical methods that do not fully accommodate the subpopulation heterogeneity.} This article proposes novel optimization criteria for calibrating and evaluating candidate statistical designs of master protocols in the presence of the potential for treatment effect heterogeneity among enrolled patient subpopulations. The framework is applied to demonstrate the statistical properties of conventional study designs when treatments offer heterogeneous benefit as well as identify optimal designs devised to monitor the potential for heterogeneity among patients with differing clinical indications using Bayesian modeling.
\end{abstract}
 \keywords{adaptive design, Bayesian analysis, \AK{hyperparameter calibration}, master protocols, multiple comparisons}

\maketitle

\section{Introduction}
\pari

Breakthroughs in cancer biology and immunology have defined new research programs emphasizing the development of non-cytotoxic therapies that target specific pathways in tumor cells or promote anti-cancer immunity. Innovations in clinical trials have followed with ``master protocols,'' that are defined by inclusive eligibility criteria and devised to interrogate multiple therapies or histologies for a given therapy \cite{WandL17, RandS16}. The use of master protocols for oncology has become more common with the desire to improve the efficiency of clinical research. The Lung-MAP study used a master protocol for a Phase II/III comparative trial designed to evaluate biomarker-matched therapies of patients previously treated with advanced squamous non-small-cell lung cancer (NSCLC) \cite{Steueretal15}. Shortly after initiation, the FDA approved a new immunotherapy for the same population. The design's flexibility facilitated subsequent modification of the standard of care, which preserved the relevance of the study \cite{WandL17}. The CREATE study master protocol evaluates the efficacy of crizotinib in patients with ALK or MET mutations and six different tumor histologies, where each histology constitutes a subtrial of ALK/MET+ or ALK/MET- tumors \cite{RandS16}. The NCI-MATCH trial uses a master protocol to enroll participants into drug-mutation-specific baskets while also incorporating genetic screening \cite{Cunananetal17}.

Trial designs for ``precision medicine'' often partition study cohorts into putatively homogeneous subtypes on the basis of a shared genetic feature or common treatment target. Basket trials, for example, are designed with the hypothesis that the presence of a molecular marker predicts response to a targeted therapy independent of tumor histology \cite{RandJ15, RandS16, Cunananetal17}. While expedient conceptually, the assumption that an individual patient's expectation of treatment benefit is well characterized by the presence of a drug target alone may be limiting in practice. Proliferation rates of tumor cells are known to vary within and between tumor types \cite{RandK14}. Moreover, contemporary sequencing methodologies have revealed extensive genetic variation between different types of cancer. Treatment benefit heterogeneity may not be well characterized on the basis of a single molecular feature. As drug developers pursue drug approvals with broad labels (e.g. Pembrolizumab whose indication was expanded to include any unresectable or metastatic solid tumors with microsatellite instability in 2017) trials will evaluate multiple disease indications for ``agnostic'' effects. For these trials, subpopulation analysis becomes central to the study design and should be considered in the design and evaluation of trial operating characteristics.

This article first presents and reviews the statistical operating characteristics that need to be explicitly considered in the design stage. We then present a novel framework for use in the design stage of trials with inclusive enrollment for which subpopulation heterogeneity is central to the study's hypothesis in order to identify the optimal design given the potential uncertainty of the trial's outcomes. The proposed criteria facilitates the consideration of multiple scenarios based on proposed and accommodates all possible trade-offs of type I error, marginal or family-wise, and power, extending to scenarios that are less extreme than those previously considered (e.g. \cite{FandK13}). The criteria are then used to illustrate how to identify an optimal basket trial design that monitors the potential for heterogeneous treatment benefit among patients with differing clinical indications based on Bayesian modeling using the multisource exchangeability model \cite{HandL18, KKandH17}.

\section{Multi-Indication Trial Statistical Considerations}
\label{sec:operatingchars}
Trials devised to enroll patients with a common drug target arising from multiple, potentially heterogeneous clinical indications pose additional complexities that require careful consideration. In particular, trialists need to decide whether to acknowledge subpopulation heterogeneity in the statistical analysis strategy as well as select an appropriate level of type I error control and discerning its implications for the design statistical power in the presence/absence of heterogeneity. The convention paradigm for trial design would suggest consideration of a single possible scenario or outcome, however this reduces the potential scope of possibilities when explicitly considering multiple indications, is impractical to select in advance, and fails to provide a comprehensive evaluation of the trial's properties across various scenarios of heterogeneity.

\subsection{Analysis Strategies}
For trials enrolling multiple indications or subpopulations, the consideration of how to incorporate subpopulations into the statistical analyses has traditionally existed on a spectrum that spans treating each indication as independent to one of naive pooling under an assumption of full statistical ``exchangeability'' of the subpopulations with one another. At the extreme of treating all indications as exchangeable and pooling them together, inference is based on the average population effect over all subpopulations and ignores the potential for any heterogeneity amongst indications. While power may be increased under this analysis strategy with the pooling of all data, in the presence of heterogeneity the indication-specific effect will not be identifiable. An example of this approach in practice was the single arm, open-label, Phase II basket trial of Imatinib in 40 different pathologic diagnoses that were hypothesized to have imatinib-sensitive kinases based upon encouraging results from an earlier study in Philadelphia chromosome positive chronic myelogenous leukemia patients \cite{KandV05,Heinrichetal08}. The study was ultimately not designed to evaluate subpopulation-specific effects in each of the 40 diagnoses, and all indications were pooled together for one analysis population.  On the other extreme of the spectrum, designs could evaluate each indication independently. This approach explicitly accommodates the potential for heterogeneity amongst indications, however it may suffer from limited power when there is low enrollment to an indication. For example, a basket trial was designed to determine the potential efficacy of vemurafenib on \emph{BRAF V600} mutations for non-melanoma patients after encouraging results were seen in melanoma patients \cite{Sosmanetal12,Hymanetal15,Cunananetal17}. However, the trial did not have equal accrual across all indications with some yielding few responses, with overall conclusions being difficult to ascertain on the basis of the trial alone \cite{HKHandL18}.

The development of statistical methods to facilitate sharing across exchangeable indications, while avoiding the sharing of information among non-exchangeable indications, provides a flexible approach relative to the strict dichotomy of analyzing the indications as independent or pooled that may increase statistical power while controlling bias \cite{BAKandC16, BCLandM10, HandL18, Thalletal03, LRandC09, leon2012borrowing}. While these modeling strategies have encouraging results and have the potential to arrive at conclusions with fewer patients, other authors have encouraged caution when considering these methods for incorporation to a proposed trial with multiple indications \cite{Cunananetal17,HandL18,FandK13}. One of the challenges noted, is that the decisions made for the analysis strategy can have profound implications on the expected enrollment distribution, the trial's operating characteristics, and the likelihood that a trial is robust to departures from the design assumptions in its ability to delineate truly effective indications.

\subsection{Operating Characteristics}
The fundamental consideration when designing trials with multiple indications is the method of type I error control and its expected strength. The potential for incorrectly identifying a null indication as effective can be summarized by marginal or family-wise type I error rates. The \textit{marginal} type I error rate estimates the type I error rate within each indication separately, whereas \textit{family-wise} type I error rates consider the entire trial in violation if a single indication falsely rejects the null hypothesis. The type I error rate that is most appropriate for a given study is partially determined by its objectives (i.e., exploratory versus confirmatory), with family-wise type I error rates reflecting more stringent control against false positives within individual indications when compared to marginal type I error control. In the case of pooling all indications together, the marginal and family-wise approaches are identical. 

An additional consideration with regards to type I error control is the strength of control across expected scenarios. In the design stage for trials with multiple indications the treatment effect size no longer exists as a one dimension parameter, but may vary within each indication. To address this challenge, Dmitrienko, Tamhane, and Bretz describe two ``senses'' of family-wise error rate control, weak versus strong, that can be beneficial in helping to frame potential design considerations \cite{DTandB09}. Weak control suggests that the family-wise type I error rate is only guaranteed for the global null scenario, where the null hypothesis is true for all indications, but may become inflated if other scenarios are encountered. \AK{In contrast, the definition of strong control would strictly require that for any configuration of null and alternative scenarios the family-wise type I error rate is maintained at the desired level. Given the challenge in evaluating this in practice and with simulation studies, we relax this definition to consider strong control so that the family-wise type I error rate is maintained across different reasonable scenarios.} For studies with multiple indications, we generalize this concept to control of marginal or family-wise type I error and consider strength of control over a continuum. Studies that only consider operating characteristics for the global null scenario by default use weak control. \AK{Alternatively, strong control of the type I error accommodates “mixed alternative” scenarios.} While many articles discuss various details of the trade-offs described above, none have provided generalized guidelines or criteria to inform the selection of optimal designs for studies with multiple indications across the spectrum of type I error control \cite{Cunananetal17,FandK13,HandL18}. \AK{Additionally, for statistical models with hyperparameter specification, there is no prior framework to identify the optimal values \textit{a priori}.}  One potential solution would be to evaluate the operating characteristics of a trial across a variety of assumptions and scenarios, however it is challenging to efficiently synthesize these potentially disparate results and one may end up implicitly averaging various scenarios to represent an overall summary of the operating characteristics. \AK{Ultimately this implicit averaging may not identify the optimal design or hyperparameter values to appropriately maintain the desired operating characteristics after the study is completed, potentially leading to a failed study.}

\section{Generalized Design Criteria}
\label{sec:designcriteria}
\pari
To address the lack of generalized guidelines for the design of trials enrolling multiple potentially heterogeneous indications, this section presents novel design criteria devised to accommodate the full spectrum of type I error control from weak to strong scenarios. To reflect the oncology setting and seminal simulation study \cite{FandK13}, we focus our discussion on single-arm trials (i.e., trials with no control group) for which response consists of a binary endpoint such as objective tumor response. \AK{To facilitate discussion of the criteria and methods presented, we first introduce some notation.} Consider a trial devised to enroll a total of $N$ patients with a common genetic mutation arising from a total of $J$ clinical indications for which it is hypothesized, but not substantiated, that the intervention provides similar effects across all subpopulations. Let $i$ index individual patients such that $Y_{i,j}=1$ indicates the occurrence of a successful response for the $i$th patient within indication $j$, while $=0$ indicates treatment failure. Let $n_j$ denote the number of patients observed with indication $j$ and denote the total number of responses in basket $j$ by $S_j$ $=$ $\sum_{i=1}^{n_j}Y_{i,j}.$ 

Consider the context of a trial enrolling $J=5$ indications such that the response rate for each indication can assume one of two values which correspond to the null and alternative hypotheses. This yields six possible scenarios which range between the global null and global alternative scenarios (Table \ref{tab:scenarios}). The global null assumes all indications are identically ineffective, and thus assume the null response rate (Scenario 1). At the opposite end of the spectrum, all indications are identically effective and assume the alternative response rate (Scenario 6). In between, four intermediate combinations exist (Scenarios 2-5), where Scenario 2 contains four ineffective indications and one effective indication whereas Scenario 5 conversely includes one ineffective indication and four effective indications.

\begin{table}[ht]
\centering
\begin{tabular}{l|ccccc}
\hline
         & \multicolumn{5}{c}{Indication} \\ \cline{2-6}
Scenario & 1   & 2   & 3   & 4   & 5  \\ \hline
1 (Global Null) & Null   & Null   & Null   & Null   & Null  \\
2        & Null   & Null   & Null   & Null   & Alt.  \\
3        & Null   & Null   & Null   & Alt.   & Alt.  \\
4        & Null   & Null   & Alt.   & Alt.   & Alt.  \\
5        & Null   & Alt.   & Alt.   & Alt.   & Alt.  \\
6 (Global Alternative) & Alt.   & Alt.   & Alt.   & Alt.   & Alt. \\ \hline
\end{tabular}
\caption{The six possible scenarios for a trial enrolling 5 indications (i.e., subpopulations), where each corresponds to either the null (Null) or alternative (Alt.) hypotheses.}
\label{tab:scenarios}
\end{table}

Traditionally, when designing a trial with multiple indications, only one of the aforementioned scenarios are chosen to summarize the power and type I error and to identify optimal hyperparameter selection. Depending on the analysis strategy, type I error, and strength of control chosen, the choice of a single scenario could present overly critical or optimistic estimates. In the case of basket trials, this traditional approach fails to reflect a reality wherein people inherently synthesize multiple scenarios over their uncertainty as to the true scenario that may be encountered when considering different modeling strategies or for determining how to optimize hyperparameters for a given model. For example, type I error rates are minimized for analysis strategies devised to share information across indications when all indications are ineffective (i.e., the global null scenario). The presence of one effective basket will increase the type I error rate when sharing information across indications. Thus, an analysis strategy may appear sub-optimal when compared to other strategies when evaluated on the basis of a single scenario. Further, despite the fact that pre-clinical evidence may suggest that a particular molecularly targeted therapy offers ``tumor agnostic'' effects, indication heterogeneity isn't predictable prior to implementation of human study. Thus, design selection based on trial operating characteristics should reflect the synthesis of many possible trial outcomes.

The methodology presented below is devised for use in the design stage of studies enrolling multiple indications to yield weighted summaries of operating characteristics for use in evaluating different modeling strategies and to identify the optimal hyperparameter specification for a given model. The approach extends optimal design selection to encompass numerous possible trial outcomes. Upon elicitation of null and alternative hypotheses, weights may be assigned to scenarios in relation to counts of indications that assume null versus alternative values. \AK{We propose an objective generalized approach based on the scenarios described in Table \ref{tab:scenarios} which considers the number of null or alternative indications per scenario in the weighting, where a trial with $J$ indications has $J+1$ scenarios.} Let $b_{X}$ denote the number of null indications (for type I error) or alternative indications (for power) in Scenario X, where X is the identifier for the scenario \AK{in Table \ref{tab:scenarios}}, and $s$ is a hyperparameter for the weighting scheme. We propose a standardized approach to defining the weight assigned Scenario X, $\omega_{X}$, as

\[ \omega_{X} = \frac{ b_{X}^{s} }{ \sum_{x=1}^{J+1} b_{x}^{s} }. \]
\AK{While the formula as presented is for the context where there are two possible outcomes, null or alternative, we note that it could be modified to consider scenarios with additional alternative outcomes or mixtures of scenarios that may be under consideration in the design stage of a study.}

\begin{figure}[ht]
\centering
\includegraphics[scale=1]{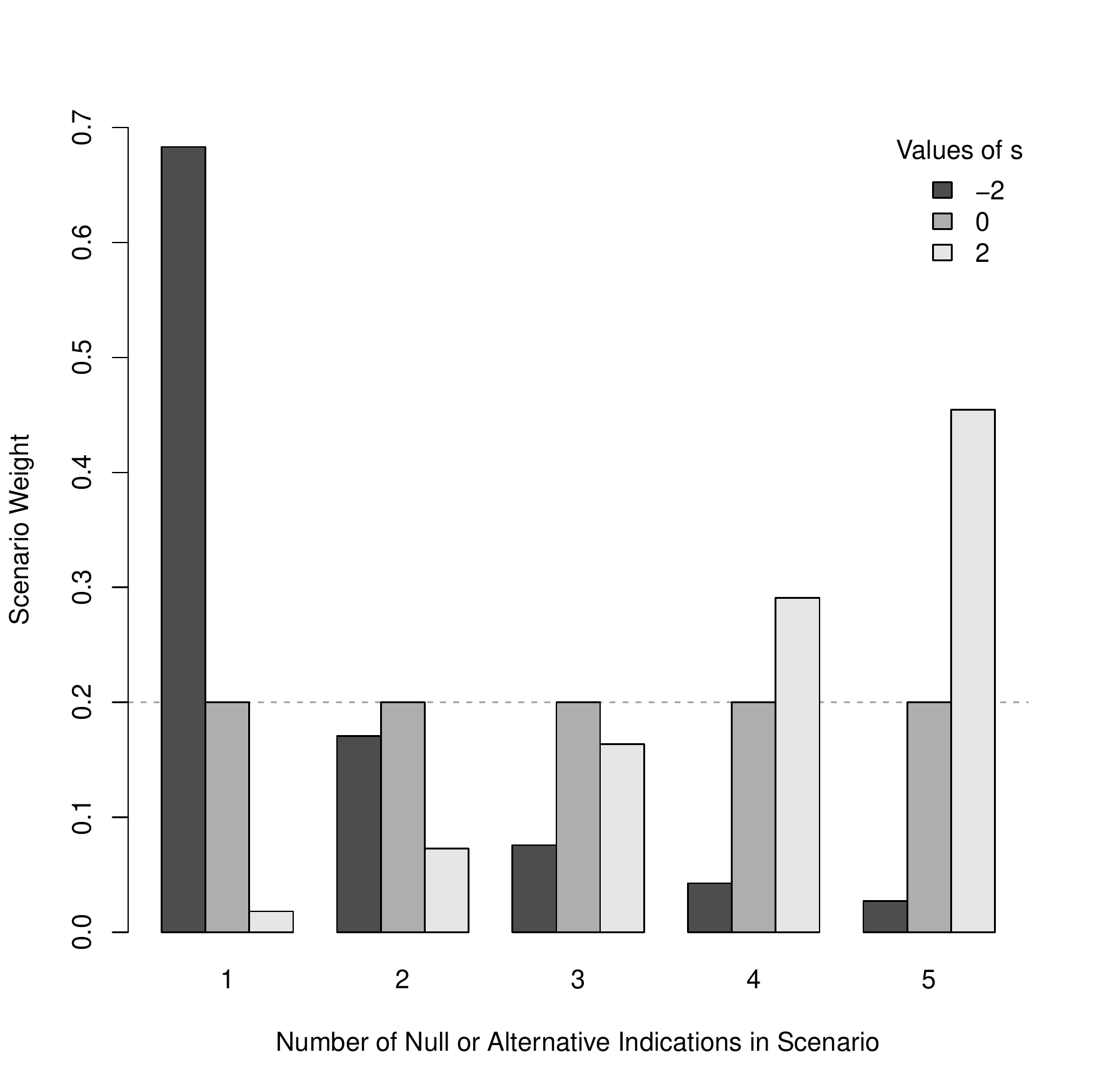}
\caption{Scenario weights for a trial enrolling five indications defined by the number of indications and various values of $s.$}
\label{fig:scenweights}
\end{figure}

The proposed weighting scheme has some \AK{appealing} properties. If $s=0$, all scenarios are weighted equally. With $s>0$, scenarios with more indications of a given criteria (null or alternative) receive greater weight, while $s<0$, increases the weight assigned to scenarios with fewer indications. Figure \ref{fig:scenweights} illustrates the scenario weighting scheme considering values of $s$ of -2, 0, and 2 for a trial enrolling five indications (Supplementary Materials Table 1 provides examples for additional values of $s$ and scenarios enrolling 10 indications). \AK{It should be noted that to avoid overly complex formulations for the proposed $\omega_{X}$, positive and negative values of $s$ are not purely symmetric about 0.} The resultant weights, however, do follow the same trend by which global scenarios are weighted more heavily as values of $s$ increase in absolute value. 

A single value of the hyperparameter $s$ may not be appropriate for synthesis of both the type I error rate and power across scenarios. For evaluating weighted power, the framework accommodates multiple values of $s$ that differentially weight alternative indications (\AK{Table \ref{tab:scenarios}} Scenarios 2-6 for 5 indications, denoted $s_{a}$) from null indications used to calculate weighted type I error (Scenarios 1-5 for 5 indications, denoted $s_{n}$). For example, in identifying the optimal hyperparameter values and posterior probability thresholds for a Bayesian model to have a weighted type I error rate less than or equal to a desired level, one may wish to optimize these parameters under scenarios with fewer null baskets (i.e., negative values of $s_{n}$), but then wish to compare the impact of these potential hyperparameters for power across all scenarios equally (i.e., $s_{a}$=0). This potential for separate specifications provides further flexibility for summarizing operating characteristics and identifying optimal hyperparameter values across a variety of potential scenarios based upon the context of the given study. 

\begin{figure}[ht]
\centering
\includegraphics[scale=1]{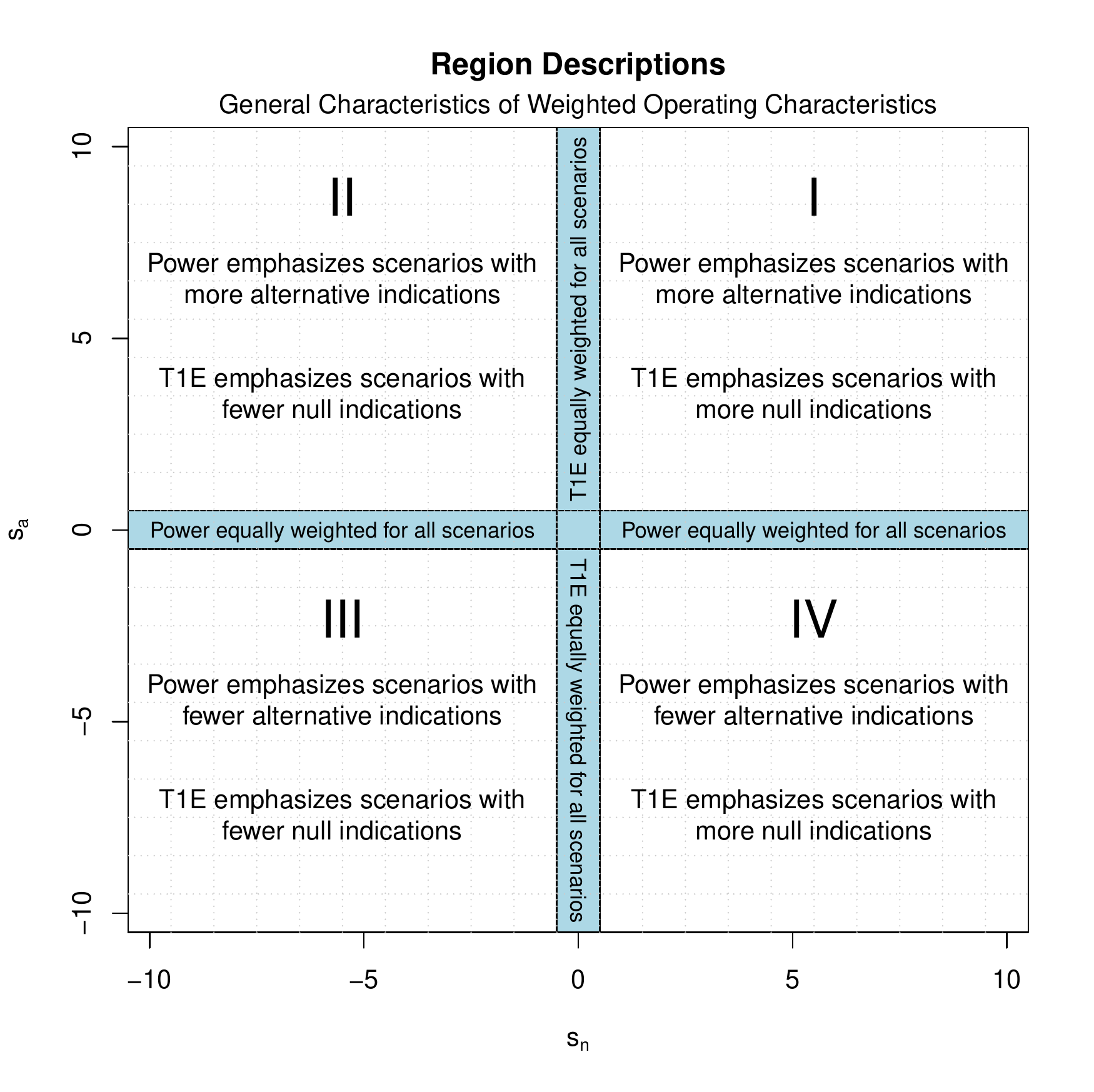}
\caption{Nine distinct regions defined by the proposed weighting scheme for combinations of $s_n$ ($s$ power used for null indications) and $s_a$ ($s$ power used for alternative indications).}
\label{fig:heatmap_description}
\end{figure}

Figure \ref{fig:heatmap_description} partitions combinations of $s_{n}$ and $s_{a}$ into a two-dimensional plane comprised of four possible quadrants. Boundaries between quadrants result from either $s_{n}=0$ or $s_{a}=0.$ Quadrant I, delineated by $s_{n}>0$ and $s_{a}>0$, depicts the region wherein analysis strategies devised to share information across indications will outperform subpopulation specific analyses. Note that large values of $s_{n},s_{a}$, yield weighted characteristics that favor the global null and alternative scenarios, which assume all indications are exchangeable. Conversely, Quadrant III, when $s_{n}<0$ and $s_{a}<0$, represents the region least amenable to sharing information because it more heavily weights the scenarios that include singleton null and alternative indications. Quadrants II and IV reflect what may be seen as more realistic settings. Quadrant II emphasizes scenarios with few null indications and more alternative indications, whereas the opposite is true for Quadrant IV. A special setting is when $s_{n}=0$ \textit{and} $s_{a}=0$, which represents the agnostic setting where all scenarios are weighted equally and may be appropriate when there is not enough evidence or consensus to identify ideal quadrants. For example, with $J=5$, if $s_{n}=-10$ and $s_{a}=0$, the type I error rate for the scenario with only one null indication receives nearly all of the weight (99.9\% weight), while power is calculated weighting all five scenarios with alternative indications equally.

With the introduction of a method to consider multiple scenarios simultaneously, we can define a sequence of steps required to identify an optimal design for any analytic strategy. 
First, the form of type I error control, marginal or family-wise, must be selected. Second, the hyperparameters $s_{a}$ and $s_{n}$ must be chosen based on the desired weighting of scenarios to be summarized for the given study in order to best evaluate the optimal design under consideration for implementation. One's choice of hyperparameters should reflect the likelihood and importance of identifying singleton indications versus the importance of describing predominate trends among all indications. Previous authors, such as Freidlin and Korn, provided recommendations for which scenarios that included a single null indication were assigned $100\%$ weight \cite{FandK13}. Choices of the hyperparameters may naturally follow in accordance with a trial's phase and intent, as providing exploratory or confirmatory evidence. The third step entails simulation studies devised to identify decision rules that achieve the desired level of weighted type I error rate control. These decision rules can be characterized by a set of posterior probability thresholds for Bayesian designs or thresholds for test statistics with frequentist designs. Having simulated a large number of replicate trials, decision thresholds are computed over a grid of potential values (i.e., posterior probabilities or test statistics). The weighted type I error can be computed for each threshold and admissible thresholds are identified as yielding weighted type I error less than or equal to the target. Weighted power is then computed for admissible decision rules. Finally, designs with various analysis strategies can be compared with respect to numerous possible trial outcomes and the optimal model (and potential hyperparameters) with a corresponding decision threshold can be selected for use in the proposed trial. In situations where multiple hyperparameters achieve the maximum observed weighted power, additional rules can be specified for which is selected for implementation (e.g., choose the smallest, largest, or average of eligible hyperparameter values). Returning to our example with $J=5$, $s_{n}=-10$ and $s_{a}=0$, the optimal hyperparameters would be selected based on the weighted type I error rate favoring scenarios with fewer null indications that maximize the power that is weighted from all scenarios equally.

\section{Illustration of Proposed Design Criteria}
\label{sec:illustration}
\pari
In order to illustrate the proposed design criteria, we first describe potential analysis strategies, including approaches representing naive pooling of data or subpopulation specific analyses which treat indications as independent units and a Bayesian design that facilitates information sharing across indications with multisource exchangeability models (MEM) \cite{KKandH17}. To facilitate direct comparisons with simulation studies previously reported by Freidlin and Korn, we present simulation results for scenarios with 5 or 10 indications \cite{FandK13}.

\subsection{Analysis Strategies}
\label{sec:mems}
\pari
In this subsection we define the analysis strategies compared in our simulation study, as well as describe the concept behind the Bayesian MEM. Lacking multivariate specification of relationships among subpopulations, conventional analysis strategies represent the extremes of either pooling all indications or analyzing each indication independently of the evidence acquired for other indications. More specifically, ``pooling'' in this context refers to an analysis strategy which is devised to evaluate evidence of effectiveness for all indications via statistical inference of a single response rate that is estimated using data combined from all indications. Our simulation study uses the exact binomial test to evaluate the statistical significance of a pooled treatment effect, where the corresponding p-value from the test is simply compared to the predefined level of significance (e.g., $\alpha=0.10$). Relying on only one test, marginal and family-wise type I error rates are identical for the pooled approach. 

At the other extreme, independent analyses infer a total of $J$ response rates estimated from the data observed within specific indications. Trials simulated with this strategy conduct $J$ exact binomial tests, one for each indication. Analysis of each indication independently could use either form of type I error control. Family-wise type I error is used to control the likelihood of a single false rejection of the null hypothesis. To control the family-wise type I error rate, we correct for multiple comparisons across the $J$ indications using a simple Bonferroni correction of $\frac{\alpha}{J}$. For example, with 5 indications and type I error control at $\alpha=0.10,$ an indication would confer significant statistical evidence to reject the null hypothesis if its exact binomial p-value is $< 0.02.$

The MEM approach induces multivariate analysis through a Bayesian hierarchical model devised to estimate the extent to which indications represent pairwise statistically exchangeable subpopulations. This implies that each indication has a potentially different measure of (pairwise) exchangeability for each other indication. Our simulation study considers the empirical Bayes (EB) and constrained-EB (cEB) priors for symmetric MEMs described by Hobbs and Landin\cite{HandL18}. Briefly, the EB identifies and conditions the analysis with respect to the MEM which maximizes the marginal data likelihood. Acknowledging the uncertainty about identifying the ``true'' MEM from the marginal likelihood, the cEB prior constrains shrinkage using an upper bound (UB) on the prior probability of inclusion for each source. The null hypothesis within each indication is rejected if the posterior probability is greater than a predefined threshold which is identified to maintain a desired type I error rate. We can further note that setting UB=0 represents a Bayesian implementation of the concept underlying the independent approach.

\subsection{Simulation Design}
\label{sec:simulations}
\pari
Design evaluation using the proposed weighted operating characteristics framework is demonstrated using simulation. We consider the context of wanting to design a fixed sample size trial for two sets of scenarios with 5 or 10 indications, where each indication is planned \textit{a priori} to accrue 25 patients. Following the simulation design of Freidlin and Korn\cite{FandK13}, null and alternative response rates assumed the values of 10\% and 30\%, respectively. For the trial with 5 indications, there are six possible scenarios which range from all indications being null (Scenario 1) to all indications being alternative (Scenario 6), and four intermediate combinations (Scenarios 2-5) (i.e., Table \ref{tab:scenarios}). The trial with 10 indications has 11 similarly defined scenarios. For each scenario, 10 000 simulated trials were completed. \AK{For each simulated trial, the MEM prior specifications detailed below and the two frequentist approaches were applied. These simulated trials were then used for the calibration of a given ($s_{n}$,$s_{a}$) to identify the optimal MEM hyperparameter and the needed posterior probability thresholds to maintain the type I error rate. Finally, the operating characteristics were summarized from the simulated trials by scenario.}

While, in practice, ($s_{n}$,$s_{a}$) should be defined in advance for the context of the specific study being planned, we illustrate the weighted operating characteristics over a grid of hyperparameters ($s_{n}$,$s_{a}$) ranging from -10 to 10 by increments of 1 to demonstrate a full range of possible conclusions under different weighting paradigms. Posterior probability thresholds for each combination were identified to maintain weighted marginal or family-wise type I error rates of 10\%. 

To illustrate the ability of the proposed framework to assist in hyperparameter selection, we consider the Bayesian symmetric MEM approach with a constrained EB prior. For this approach we consider UB hyperparameters over 29 values ranging from UB=0 (no borrowing) to UB=1 (all weight to the MEM maximizing the integrated likelihood). The optimal UB parameter is then selected to maximize power for each combination of $s_n$ and $s_a$ among admissible models for the specified extent of type I error control. When multiple UB parameters maximized power, the smallest value was chosen (e.g., if UB=0.05 and UB=0.10 both maximized power, UB=0.05 was selected as optimal).

We first present a comparison for the scenarios with 5 indications to demonstrate the improvement in calibration between the traditional approach of selecting one scenario to optimize hyperparameter values on versus the proposed weighted operating characteristics approach for the agnostic weighting where $s_{n}=s_{a}=0$. We then present the results for weighting operating characteristics for the frequentist independence and pooling approaches and the Bayesian symmetric MEM approach across the grid of ($s_{n}$,$s_{a}$) values. With the previously described approach to calibration to maintain 10\% type I error control, operating characteristics for the Bayesian MEM models based on optimal UB parameters were compared directly to frequentist independence and pooling approaches. Simulations were conducted in R \cite{Rref}.

\subsection{\AK{Hyperparameter} Calibration for Optimal Models}
\pari
In this section, we examine the different conclusions for \AK{hyperparameter} calibration in our simulation study with 5 indications if the trial is optimized for only one scenario as compared to the proposed weighted operating characteristics approach. This means that \AK{during the design phase} the optimal UB hyperparameter for the MEM assuming a constrained EB prior is calibrated either for only one scenario versus the proposed framework which utilizes all scenarios in calibrating the parameters. \AK{In either approach, the selected hyperparameter would be used for the analysis after trial completion.} For our initial step, Table \ref{tab:optimal_ub} represents the values of UB that maximize the power for each of the separate scenarios including at least one null indication, with many scenarios having a range of UB parameters that may be considered optimal with respect to power. Recall, we conservatively specify the lowest possible UB value is selected as optimal when multiple are eligible. We can see that only examining the global scenarios would result in aggressive borrowing with UB=1, whereas the intermediate scenarios are more conservative with UB$\leq$0.12.

\begin{table}[ht]
\centering
\begin{tabular}{l|c|c}
\hline
Scenario (\# of null, & \multicolumn{2}{c}{Range of Optimal UB Hyperparameters} \\ \cline{2-3}
alternative indications) & Marginal Type I Error & Family-wise Type I Error \\ \hline
1 (5 null, 0 alternative)                      & 1.00                  & 1.00                     \\
2 (4 null, 1 alternative)                      & 0.05-0.09             & 0.03-0.12                \\
3 (3 null, 2 alternative)                      & 0.01-0.08             & 0.03                     \\
4 (2 null, 3 alternative)                      & 0.03-0.08             & 0.10                     \\
5 (1 null, 4 alternative)                      & 0.00-0.10             & 0.00-0.10               \\ \hdashline
Weighted OC ($s_{n}=s_{a}=0$) & 0.08 & 0.09 \\ \hline
\end{tabular}
\caption{The range of values for the UB hyperparameter that achieve maximum power for the given scenario calibrated to maintain the marginal or family-wise type I error rate at $\leq$10\%.}
\label{tab:optimal_ub}
\end{table}

The estimated marginal and family-wise calibrations and type I error rates, with accompanying power, are presented in Tables \ref{tab:marg_optimal} and \ref{tab:fw_optimal}, respectively, with the identified posterior probability thresholds as determined for the UB hyperparameter. These tables illustrate a few important messages. First, even when identical UB parameters are selected, different posterior probability thresholds are needed to maintain the desired type I error rate (e.g., Table \ref{tab:fw_optimal} for the rows representing Scenarios 2 and 3 with UB=0.03). This has implications for the observed operating characteristics, since the posterior probability threshold directly affects how strongly type I error is controlled as well as its corresponding power. Second, simply averaging over the optimal UB or posterior probability threshold values is not equivalent to the proposed weighted operating characteristic calibration. For example, the optimal UB for $s_{n}=s_{a}=0$ (equal weights among all scenarios) is UB=0.08 for marginal type I error rate control. Based on Table \ref{tab:marg_optimal}, however, the naive average would be 0.218, a value of UB that will more aggressively share information amongst indications than the optimal choices for 4 of the 5 scenarios. Third, the weighted operating characteristics (WOC) in Tables \ref{tab:marg_optimal} and \ref{tab:fw_optimal} are generally more consistent across scenarios when equally weighting each scenario with the proposed methodology, whereas any standalone scenario tends to have greater variability in statistical power and type I error. The findings demonstrate that calibrations that integrate multiple scenarios provide an improvement with respect to the current standard of calibrating based upon one scenario. Optimization with respect to weighted operating characteristics additionally improves upon implicit averaging across scenarios, as the scenario-wise statistical summaries fail to synthesize the myriad possible trial outcomes. Note, that statistical properties for the overall weighted summary, $s_{n}=s_{a}=0,$ are described in further detail in the next section.

\begin{table}[ht]
\centering
\resizebox{\textwidth}{!}{%
\begin{tabular}{ll|cc|cc|cc|cc|cc|cc}
  \hline
\multicolumn{2}{l}{\textit{Optimal UB for Scenario}} & \multicolumn{2}{|c}{\textit{Scenario 1}} & \multicolumn{2}{|c}{\textit{Scenario 2}} & \multicolumn{2}{|c}{\textit{Scenario 3}} & \multicolumn{2}{|c}{\textit{Scenario 4}} & \multicolumn{2}{|c}{\textit{Scenario 5}} & \multicolumn{2}{|c}{\textit{Scenario 6}} \\
Scen. & UB (PP Thresh.) & T1E & Pwr & T1E & Pwr & T1E & Pwr & T1E & Pwr & T1E & Pwr & T1E & Pwr \\ 
  \hline
1 & UB=1.00 (0.907) & 0.098 & - & 0.221 & 0.794 & 0.358 & 0.884 & 0.460 & 0.943 & 0.531 & 0.973 & - & 0.985 \\ 
  2 & UB=0.05 (0.858) & 0.096 & - & 0.100 & 0.913 & 0.120 & 0.915 & 0.175 & 0.922 & 0.220 & 0.942 & - & 0.959 \\ 
  3 & UB=0.01 (0.848) & 0.095 & - & 0.096 & 0.912 & 0.098 & 0.913 & 0.116 & 0.914 & 0.188 & 0.919 & - & 0.947 \\ 
  4 & UB=0.03 (0.860) & 0.095 & - & 0.095 & 0.912 & 0.095 & 0.912 & 0.100 & 0.912 & 0.164 & 0.913 & - & 0.938 \\ 
  5 & UB=0.00 (0.844) & 0.095 & - & 0.095 & 0.912 & 0.095 & 0.912 & 0.094 & 0.912 & 0.094 & 0.911 & - & 0.911 \\ \hdashline
  WOC & UB=0.08 (0.887) & 0.094 & - & 0.095 & 0.910 & 0.095 & 0.912 & 0.095 & 0.912 & 0.121 & 0.911 & - & 0.920 \\ 
   \hline
\end{tabular}
}
\caption{Minimum optimal UB for \textit{marginal} type I error control from Table \ref{tab:optimal_ub} and posterior probability threshold (PP Thresh.), with resulting \textit{marginal} type I error rates and power presented for all scenarios, including the weighted operating characteristic (WOC) scenario where $s_{n}=s_{a}=0$.}
\label{tab:marg_optimal}
\end{table}

\begin{table}[ht]
\centering
\resizebox{\textwidth}{!}{%
\begin{tabular}{ll|cc|cc|cc|cc|cc|cc}
  \hline
\multicolumn{2}{l}{\textit{Optimal UB for Scenario}} & \multicolumn{2}{|c}{\textit{Scenario 1}} & \multicolumn{2}{|c}{\textit{Scenario 2}} & \multicolumn{2}{|c}{\textit{Scenario 3}} & \multicolumn{2}{|c}{\textit{Scenario 4}} & \multicolumn{2}{|c}{\textit{Scenario 5}} & \multicolumn{2}{|c}{\textit{Scenario 6}} \\
Scen. & UB (PP Thresh.) & T1E & Pwr & T1E & Pwr & T1E & Pwr & T1E & Pwr & T1E & Pwr & T1E & Pwr \\ 
  \hline
1 & UB=1.00 (0.968) & 0.090 & - & 0.216 & 0.725 & 0.394 & 0.839 & 0.532 & 0.924 & 0.507 & 0.966 & - & 0.982 \\ 
  2 & UB=0.03 (0.979) & 0.096 & - & 0.100 & 0.736 & 0.088 & 0.771 & 0.064 & 0.797 & 0.032 & 0.807 & - & 0.809 \\ 
  3 & UB=0.03 (0.946) & 0.152 & - & 0.122 & 0.806 & 0.100 & 0.811 & 0.123 & 0.819 & 0.085 & 0.861 & - & 0.898 \\ 
  4 & UB=0.10 (0.956) & 0.135 & - & 0.121 & 0.788 & 0.094 & 0.808 & 0.099 & 0.814 & 0.079 & 0.843 & - & 0.889 \\ 
  5 & UB=0.00 (0.844) & 0.396 & - & 0.331 & 0.912 & 0.258 & 0.912 & 0.179 & 0.912 & 0.094 & 0.911 & - & 0.911 \\ \hdashline 
  WOC & UB=0.09 (0.957) & 0.140 & - & 0.121 & 0.796 & 0.094 & 0.809 & 0.070 & 0.811 & 0.071 & 0.818 & - & 0.875 \\ 
   \hline
\end{tabular}
}
\caption{Minimum optimal UB for \textit{family-wise} type I error control from Table \ref{tab:optimal_ub} and posterior probability threshold (PP Thresh.), with resulting \textit{family-wise} type I error rates and power presented for all scenarios, including the weighted operating characteristic (WOC) scenario where $s_{n}=s_{a}=0$.}
\label{tab:fw_optimal}
\end{table}

\subsection{Simulation Results}
\begin{figure}[ht]
\centering
\includegraphics[scale=1]{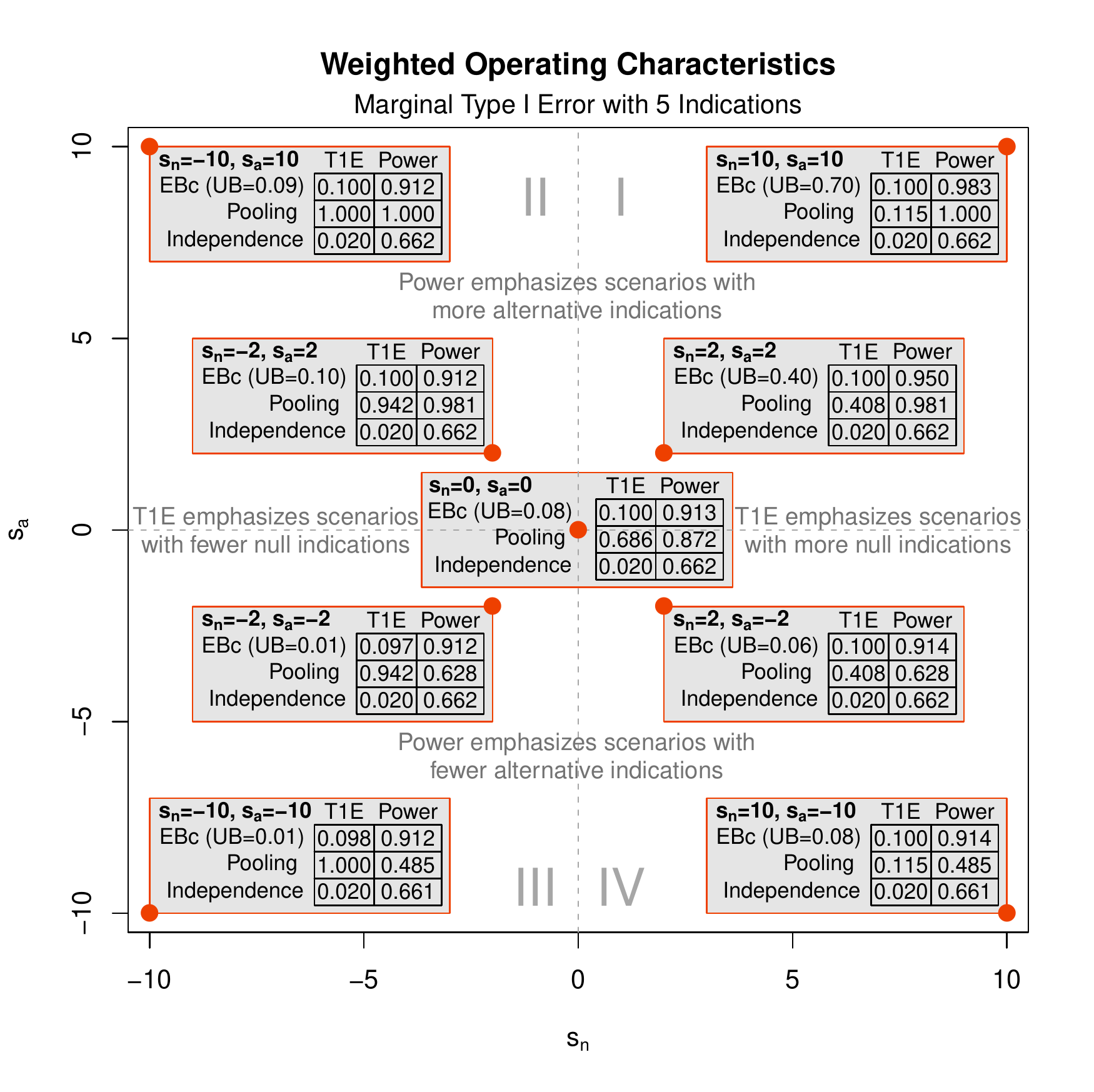}
\caption{Graphical summary of weighted operating characteristics at various combinations of $s_{a}$ and $s_{n}$ for optimal UB, pooling, and independence approaches for the 5 indication trial with weighted \textbf{marginal} type I error less than or equal to 10\%.}
\label{fig:optub_marg_5bask}
\end{figure}

\pari
Figure \ref{fig:optub_marg_5bask} presents weighted marginal type I error rates and power for various $s_{n}$ and $s_{a}$ combinations for a trial enrolling 5 indications. This figure elucidates several important aspects of the competing analysis strategies. First, independence, which treats each indication separately, results in a constant type I error rate of 0.10 but has an inadequate power of 0.66 across the spectrum of scenarios, which is 73\% of the target power of 0.90. Second, pooling results in type I error rates that are inflated to unacceptable levels across most of the spectrum of $s_n$ values. For instance, in the agnostic scenario where $s_n$ and $s_a$ are both 0 (i.e., equally weighting all scenarios), pooling demonstrates undesirable performance with a 0.686 type I error rate. When $s_{n}=2$, which weights the global null scenario with nearly 50\% weight (see Figure \ref{fig:scenweights}), the type I error rate is still grossly inflated at 0.408. In general, only if $s_n > 10$, which provides nearly all weight to the global null scenario, is the type I error rate maintained a the desired 0.10 level.

The Bayesian MEM model results in calibrations that both control type I error at the target level of 0.10 and yield power exceeding 0.912 for any combination of $s_{n}$ and $s_{a}$. In fact, at no combination does either pooling or independence both control the type I error \textit{and} improve power relative to the MEM model. Moreover, failing to share information among indications with implementation of the Bayesian MEM assuming UB=0 failed to yield optimal performance for any scenario. The top right quadrant favors larger values of UB which more aggressively share information across indications, resulting in power as high as 0.983 while maintain a 0.10 type I error rate. The other three quadrants where at least one value of $s$ is less than 0 (i.e., favoring scenarios with fewer indications) favors less borrowing (UB values $\le 0.1$), but maintains power $>0.90$. In fact, power is greater than 0.90 over all combinations of $s_{n}$ and $s_{a}$ for the MEM model, which, as presented previously, is not evident for conventional frequentist analyses.

\begin{figure}[ht]
\centering
\includegraphics[scale=1]{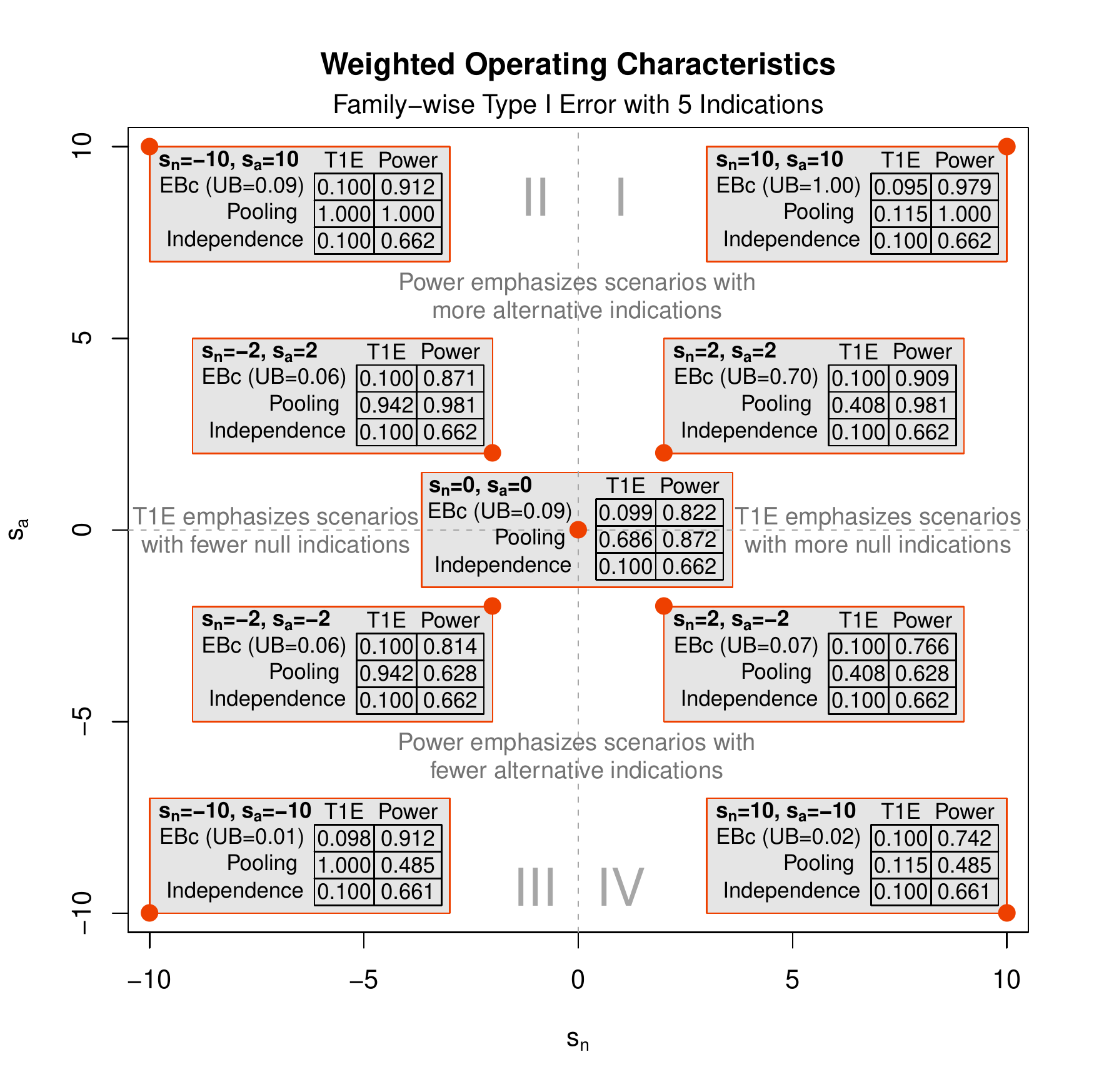}
\caption{Graphical summary of weighted operating characteristics at various combinations of $s_{a}$ and $s_{n}$ for optimal UB, pooling, and independence approaches for the 5 indication trial with weighted \textbf{family-wise} type I error less than or equal to 10\%.}
\label{fig:optub_fw_5bask}
\end{figure}

Figure \ref{fig:optub_fw_5bask} presents results for a selection of $s_{n}$ and $s_{a}$ combinations in consideration of weighted family-wise type I error rate control for a trial enrolling 5 indications. With family-wise type I error control, the Bayesian MEM model offers calibrations that control the type I error at 0.10 with power that is at worst 0.742, but as high as 0.979. In general, results are similar to the marginal type I error rate where it can be noted that it is never advantageous to use the pooling or independence approaches owing to inflated family-wise type I error rates or reduced power, respectively. Scenarios similar to the strong false-positive control of Freidlin and Korn\cite{FandK13} can be identified when $s_{n}=-10$, where almost all of the weight is given to the scenario with one null indication.

\AK{The results presented in this subsection reflect the weighted operating characteristics for the combinations of $s_{n}$ and $s_{a}$ for the optimal MEM model, the independence approach, and the pooling approach. Further, we note that a subset of results from the previous subsection in Tables \ref{tab:marg_optimal} and \ref{tab:fw_optimal} can be used to evaluate the performance of the proposed MEM model at $s_{n}=s_{a}=0$ (WOC) with the Bayesian equivalent of the independence approach for Scenario 5 where UB=0 across all scenarios. While the operating characteristics in Table \ref{tab:marg_optimal} for the marginal type I error rate comparing Scenario 5 and WOC are largely similar, the family-wise type I error rate is controlled more strongly in Table \ref{tab:fw_optimal} for WOC across scenarios versus the UB=0 in Scenario 5 with only slightly lower power in the global alternative scenario (Scenario 6). While the naive pooling approach does not have a Bayesian equivalent in Tables \ref{tab:marg_optimal} and \ref{tab:fw_optimal}, it can be noted that Kaizer, et al., demonstrated previously that it results in drastic inflation to the family-wise type I error rate for basket trials where the scenario is not the global null \cite{kaizer2019basket}.}

The simulation results for the trial with 10 indications are presented in the Supplementary Materials (Supplementary Figures 1-2 for weighted marginal and family-wise type I error rates, respectively). Results with 10 indications demonstrate similar performance to trends observed for 5 indications. Notably, sharing information among indications using the Bayesian MEM model is nearly always advantageous when compared to frequentist analysis with pooling and independence. For Quadrant IV, however, (which emphasizes identification of singleton indications) only marginal improvements in weighted power were evident for the Bayesian hierarchical model. Additional simulation results are available in the Supplementary Materials which summarize all explored combinations of $s_{n}$ and $s_{a}$ in the form of heatmaps (Supplementary Figures 3-6). 

\section{Discussion}
\label{sec:discussion}
\pari
Subpopulation heterogeneity is intrinsic to evaluations of cancer therapies emerging within the evolving oncology landscape. Consequently, trials require explicit designs that take into account the potential for subpopulation effects. When implemented with adaptive and ``seamless'' decision rules, such master protocol designs with inclusive eligibility may conserve patient and financial resources when compared to the expense of conducting a series of standalone trials for each indication \cite{Lacombeetal14,10.1093/jnci/djy196}. Conventional approaches for identifying optimal trial designs are limited by the emphasis of single, often extreme, scenarios, which may not reflect the desire or reality of all planned trials. \AK{The focus on these extreme scenarios may be, in part, motivated by existing guidance documents from regulatory authorities that consider contexts for strong type I error control under trials with multiple endpoints rather than newer master protocol designs \cite{FDA2017multiendpoints,EMA2017multiendpoints,EMA1998ICHE9}. In the context of basket trial designs there is a need to revisit the strict definition of strong type I error control in the presence of multiple indications with considerations for how to design trials that accommodate a range of possible outcomes while utilizing advances in statistical methodology that facilitate sharing information across indications. This lack of existing guidance specific to master protocols has created uncertainty in what study designs and statistical analysis strategies may be most efficient or appropriate.} 

\AK{This article presented a framework for designing optimal studies and selecting hyperparameter specification for statistical models for studies with multiple, potentially heterogeneous, indications. The framework facilitates consideration of either strong and weak type I error control with respect to outcome scenarios that are deemed reasonable during the design stage of a trial, where the strength of control can be controlled by the weighting of the scenarios.}  The resultant weighted operating characteristics and design approach synthesize numerous possible trial outcomes in order to more holistically identify the optimal design among those under consideration for the proposed trial. This has some parallels with certain methods for ensemble and machine learning that attempt to account for uncertainty with averaging over multiple scenarios. The proposed framework is general, and can be applied with consideration of marginal or family-wise control of the type I error rate. Moreover, the approach accommodates for calibration of both Bayesian and frequentist designs. 

Using conventional design criteria, authors have suggested that statistical models devised to share information across subpopulations provide minimal benefit or are detrimental in the presence of heterogeneous indications \cite{FandK13}. Their results were presented for calibrations of hierarchical models involving strong control of the type I error rate in the context of scenarios presenting singleton null indications combined with effective therapies for all other indications. When compared to an indication specific analysis, this calibration strategy yielded both reductions in the type I error rate and power for the Bayesian models, making overall conclusions difficult to ascertain. By way of contrast, our findings, which have extended the domain of evaluation beyond singleton null scenarios, suggest that it is advantageous to share information across indications for all combinations of $s_n$ and $s_a$ when implemented with the MEM modeling framework, which has demonstrated superiority with respect to other hierarchical modeling strategies \cite{KKandH17,HandL18}.  When compared to frequentist analyses that either pool indications or assume independence, at least marginal, and sometimes substantial, improvements in the weighted power were evident for the Bayesian MEM approach. In fact, with inclusive scenario evaluations and weighted type I error control, our evaluations suggest it is never considered disadvantageous to leverage the Bayesian MEM approach.

There are important limitations to note. While we attempted to propose a simple formula for determining scenario weights, our weighting scheme will always favor one of the extremes of a singleton indication, except when $s_{n}=0$ or $s_{a}=0$. The singleton scenario results in a more conservative setting where sharing information may be more challenging, whereas the global null/alternative scenario represents the ideal setting for borrowing across indications. While it seems advantageous to simplify the summary of an entire set of scenarios to a single set of weighted operating characteristics, this may only represent a marginal improvement in the clarity of our approach relative to considering separate operating characteristics for each scenario. However, even when presented with the results for more multiple scenarios, collaborators may still try to synthesize the results through informal weighting for a single takeaway, and model-specific hyperparameters may not necessarily be optimal by naively averaging over different scenarios. Our simulation considered a single treatment across all subpopulations, representing an easier context than may be encountered in designing complex real world trials. Another challenge is that use of statistical approaches to facilitate information sharing may be seen as too complex to incorporate without specialized knowledge. Recently, new software tools, such as the basket package in R to implement the symmetric MEM approach, have been created to facilitate use of these methods in future studies with multiple indications \cite{basketpackage}. The proposed framework for identifying optimal trial designs with symmetric MEMs is being extended to sequential trials with multiple indications as a future research aim. With the increasing emphasis on precision medicine, master protocols offer a pivotal tool for characterizing  treatment benefit heterogeneity. While the context of each individual trial is unique, further guidance pertaining to the statistical designs of such trials is needed from regulating agencies to promote best practices for each stage of the research process.

\begin{dci}
The last author receives research funds from Amgen, and serves as an advisor to Presagia.
\end{dci}

\begin{funding}
The author(s) disclosed receipt of the following financial support for the research, authorship, and/or publication of this article: This work was supported by the National Institute on Drug Abuse [grant number NIH R01-DA046320]; and the Case Comprehensive Cancer Center [grant number P30 CA043703].
\end{funding}

\bibliographystyle{SageV}

\end{document}


\thispagestyle{empty}
\doublespacing

\runninghead{Kaizer et al.}

\title{Supplemental materials for statistical design considerations for trials that study multiple indications}

\author{Alexander M. Kaizer\affilnum{1}, Joseph S. Koopmeiners\affilnum{2}, Nan Chen\affilnum{3}, Brian P. Hobbs\affilnum{4} }

\affiliation{\affilnum{1}Department of Biostatistics and Informatics,  University of Colorado-Anschutz Medical Campus, Aurora, CO\\
\affilnum{2}Division of Biostatistics, University of Minnesota, Minneapolis, Minnesota\\
\affilnum{3}Department of Biostatistics,  University of Texas M.D. Anderson Cancer Center, Houston, TX\\
\affilnum{4}Quantitative Health Sciences; Taussig Cancer Institute, Cleveland Clinic, Cleveland, OH
}

\maketitle

\section{Table of scenario weights}

Table \ref{tab:priorweight} provides examples of the scenario weights for various values of $s$ for both 5 and 10 indications.

\begin{table}[ht]
\centering
\begin{tabular}{r|ccccc|ccccc}
  \hline
Number of Null or Alt.  & \multicolumn{5}{c|}{Values of $s$ with 5 Indications} & \multicolumn{5}{c}{Values of $s$ with 10 Indications}  \\ \cline{2-11}
Indications in Scenario & -10 & -2 & 0 & 2 & 10 & -10 & -2 & 0 & 2 & 10 \\ 
  \hline
10 indications & - & - & - & - & - & 0.000 & 0.006 & 0.100 & 0.260 & 0.670 \\ 
  9 indications & - & - & - & - & - & 0.000 & 0.008 & 0.100 & 0.210 & 0.234 \\ 
  8 indications & - & - & - & - & - & 0.000 & 0.010 & 0.100 & 0.166 & 0.072 \\ 
  7 indications & - & - & - & - & - & 0.000 & 0.013 & 0.100 & 0.127 & 0.019 \\ 
  6 indications & - & - & - & - & - & 0.000 & 0.018 & 0.100 & 0.094 & 0.004 \\ 
  5 indications & 0.000 & 0.027 & 0.200 & 0.455 & 0.898 & 0.000 & 0.026 & 0.100 & 0.065 & 0.001 \\ 
  4 indications & 0.000 & 0.043 & 0.200 & 0.291 & 0.096 & 0.000 & 0.040 & 0.100 & 0.042 & 0.000 \\ 
  3 indications & 0.000 & 0.076 & 0.200 & 0.164 & 0.005 & 0.000 & 0.072 & 0.100 & 0.023 & 0.000 \\ 
  2 indications & 0.001 & 0.171 & 0.200 & 0.073 & 0.000 & 0.001 & 0.161 & 0.100 & 0.010 & 0.000 \\ 
  1 indication & 0.999 & 0.683 & 0.200 & 0.018 & 0.000 & 0.999 & 0.645 & 0.100 & 0.003 & 0.000 \\ 
   \hline
\end{tabular}
\caption{Scenario weights with a given number of indications meeting the criteria for various values of $s$ for trials with 5 and 10 indications considered.}
\label{tab:priorweight}
\end{table}

\newpage
\section{10 indication trial figures}
\begin{figure}[!ht]
\includegraphics[scale=1]{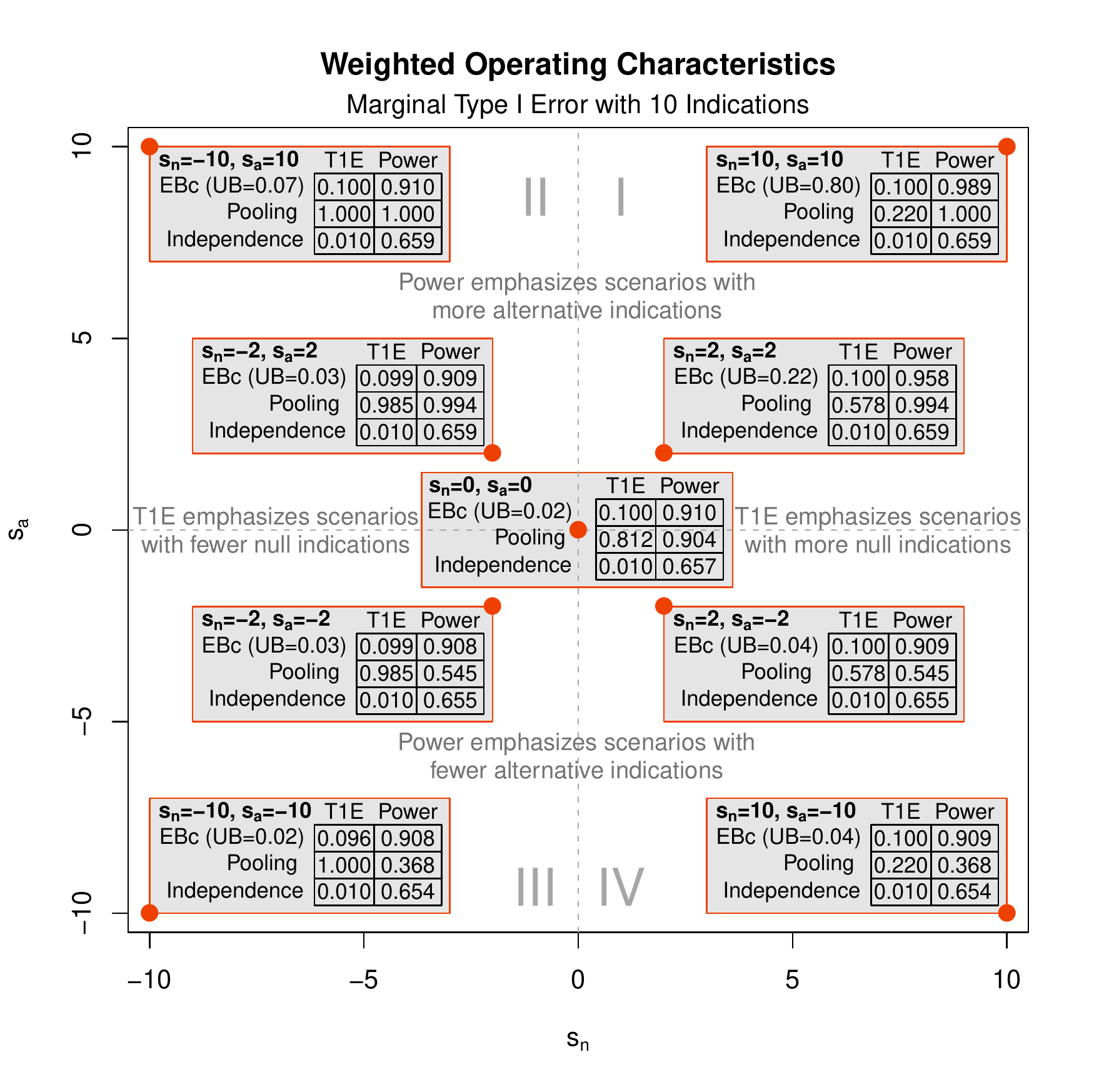}
\caption{Graphical version of weighted operating characteristics at various combinations of $s$ for optimal UB, pooling, and independence approaches for the 10 indication trial with \textbf{marginal type I error} at 10\%.}
\label{fig:optub_marg_10bask}
\end{figure}

\begin{figure}[!ht]
\includegraphics[scale=1]{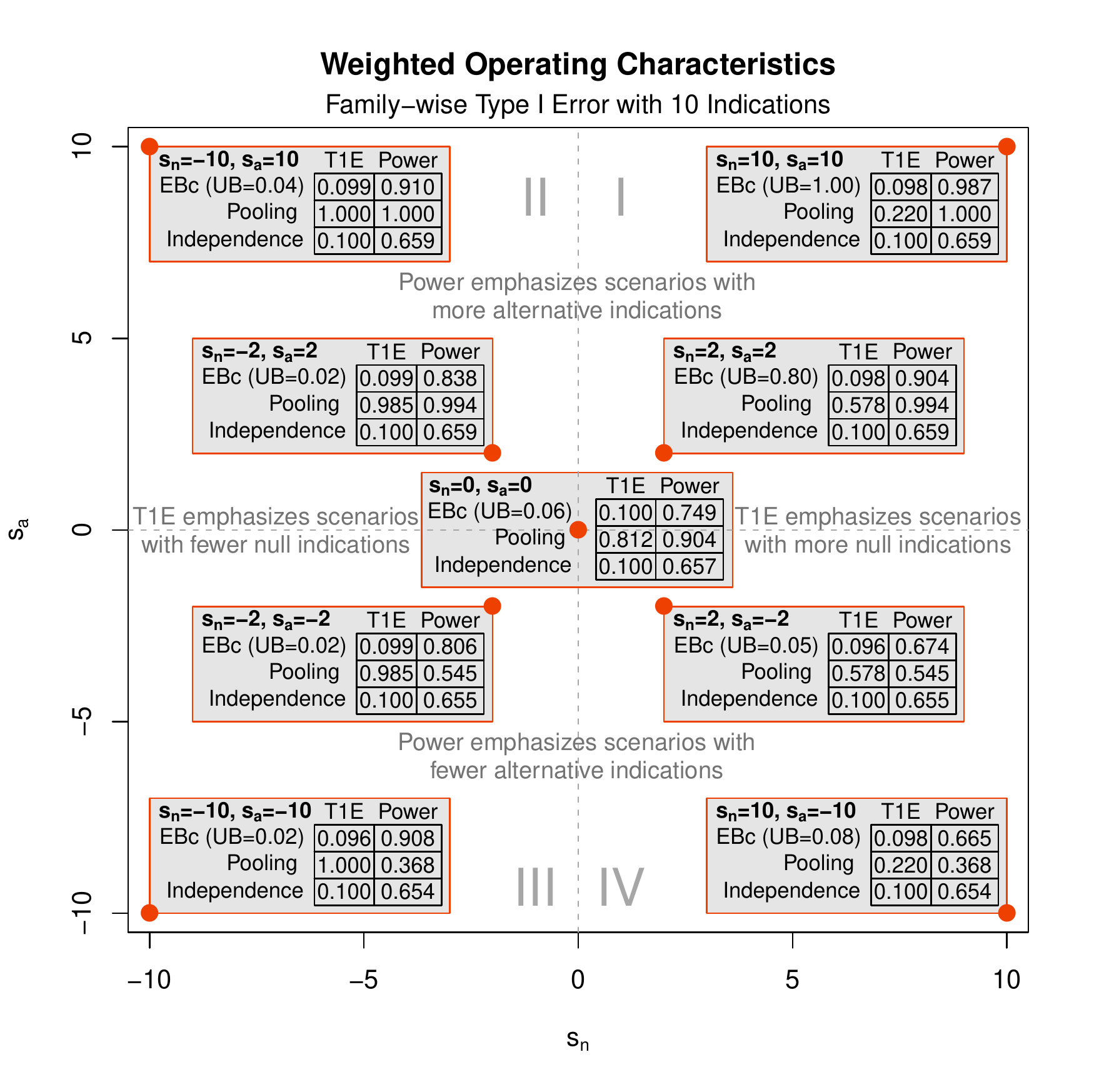}
\caption{Graphical version of weighted operating characteristics at various combinations of $s$ for optimal UB, pooling, and independence approaches for the 10 indication trial with \textbf{family-wise type I error} at 10\%.}
\label{fig:optub_fw_10bask}
\end{figure}

Figures \ref{fig:optub_marg_10bask} and \ref{fig:optub_fw_10bask} provide the weighted operating characteristics for the 10 indication trial scenario for marginal and family-wise type I error rates, respectively. As noted in the manuscript, the results a similar to the 5 indication trial results.

\clearpage
\section{Heat maps for 5 and 10 indication trials}

\begin{figure}[ht]
\includegraphics[scale=1]{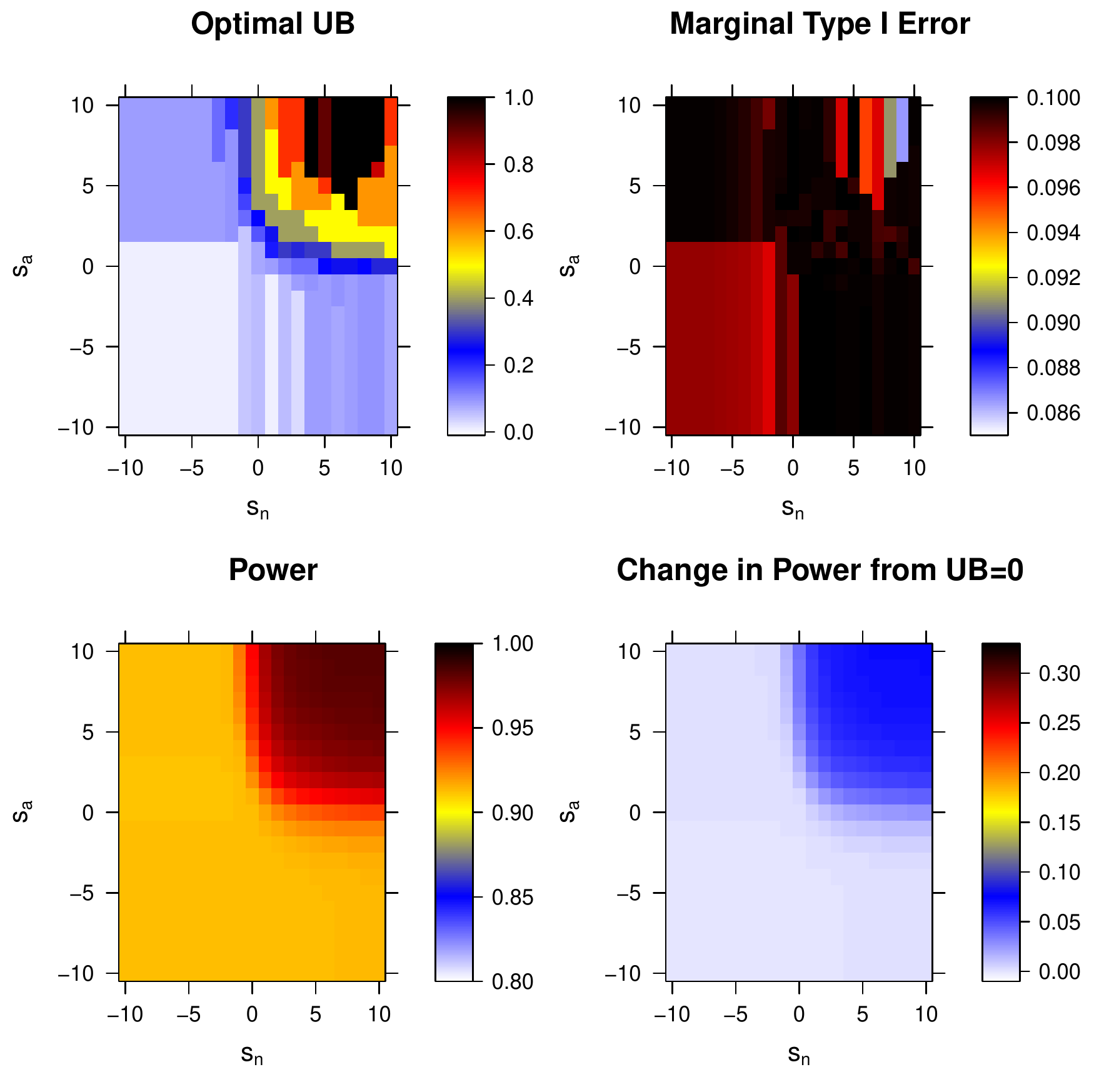}
\caption{Heat map of the optimal value of UB for 5 indication trial to maintain a weighted \textit{marginal} type I error less than or equal to 10\% for a given combination $s_{n}$ ($s$ for null indications) and $s_{a}$ ($s$ for alternative indications). Heat maps for weighted marginal type I error, weighted power, and change in power for the optimal UB compared to UB=0 (no borrowing).}
\label{fig:heatmap_2x2_marg_5bask}
\end{figure}

Figure \ref{fig:heatmap_2x2_marg_5bask} presents heat maps for the 5 indication trial over the grid of $s_{n}$ and $s_{a}$ values for the optimal UB hyperparameter which maximizes the weighted power while maintaining a weighted marginal type I error rate, the corresponding weighted marginal type I error rate and power for the optimal design, and the change in power relative to UB=0. The heat maps provide many insights. First, at no combination of $s_{n}$ and $s_{a}$ is it optimal to not borrow (UB=0), however only the top right quadrant favors larger values of UB which more aggressively borrow information across indications, whereas the bottom left quadrant favors UB values less than or equal to 0.05 which suggest minimal borrowing. Second, power is greater than 90\% over all combinations, and approaches 100\% towards the top right where increasing values of $s_{n}$ and $s_{a}$ more heavily weight the scenarios with more null and alternative indications, respectively. 

\begin{figure}[ht]
\includegraphics[scale=1]{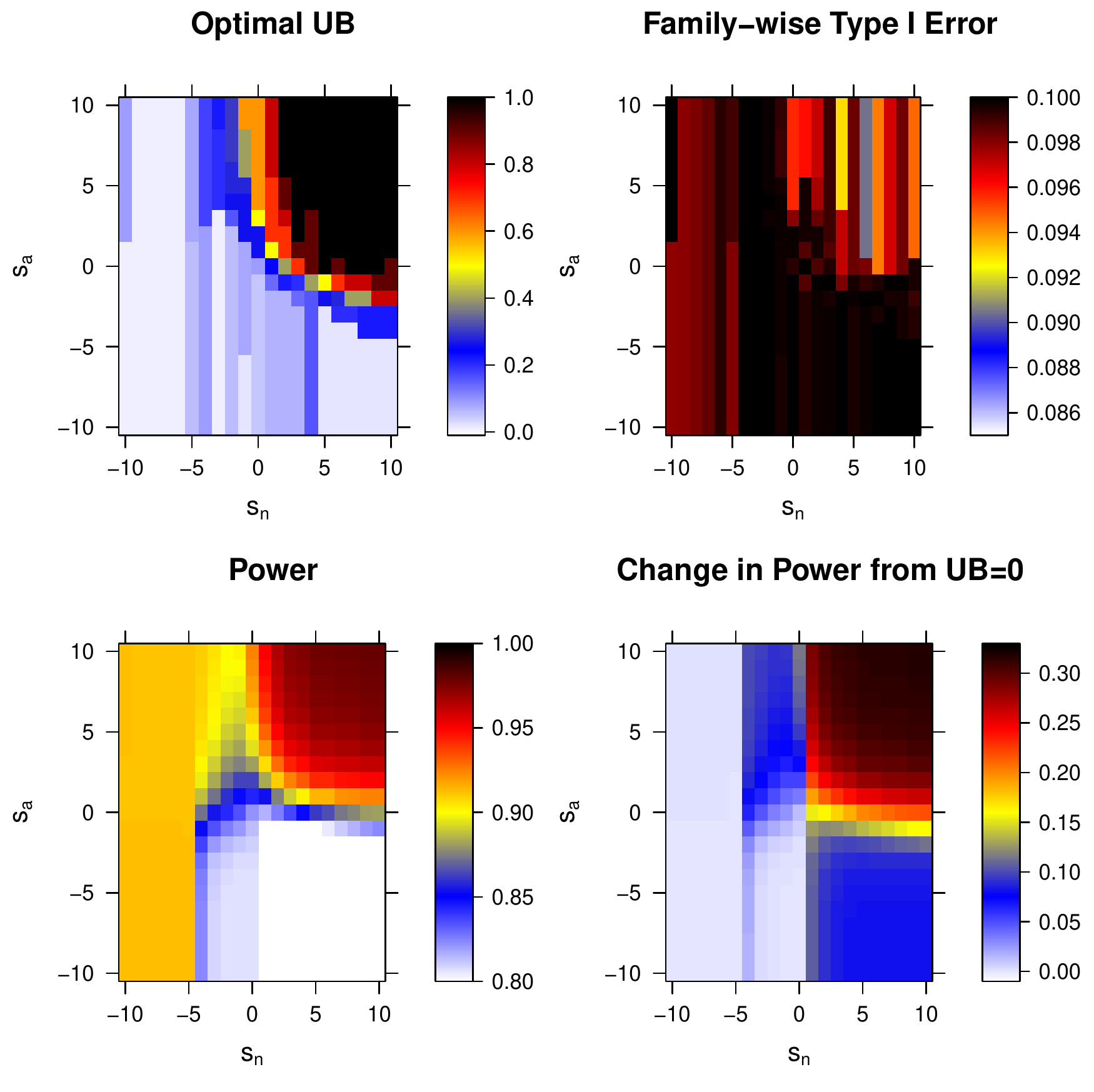}
\caption{Heat map of the optimal value of UB for 5 indication trial to maintain a weighted \textit{family-wise} type I error less than or equal to 10\% for a given combination $s_{n}$ ($s$ for null indications) and $s_{a}$ ($s$ for alternative indications). Heat maps for weighted family-wise type I error, weighted power, and change in power for the optimal UB compared to UB=0 (no borrowing).}
\label{fig:heatmap_2x2_fw_lte_5bask}
\end{figure}

Figure \ref{fig:heatmap_2x2_fw_lte_5bask} presents the same results for the 5 indication trial with the weighted family-wise type I error rate maintained at less than or equal to 10\%. In general the results are similar to Figure \ref{fig:heatmap_2x2_marg_5bask}, with UB greater than 0, in all cases, indicating that it is always optimal to borrow. The improvement in power by borrowing relative to UB=0 is more extreme with family-wise error control, as indicated by the top right quadrant approaching an increase in power of 0.30.

\begin{figure}[ht]
\includegraphics[scale=1]{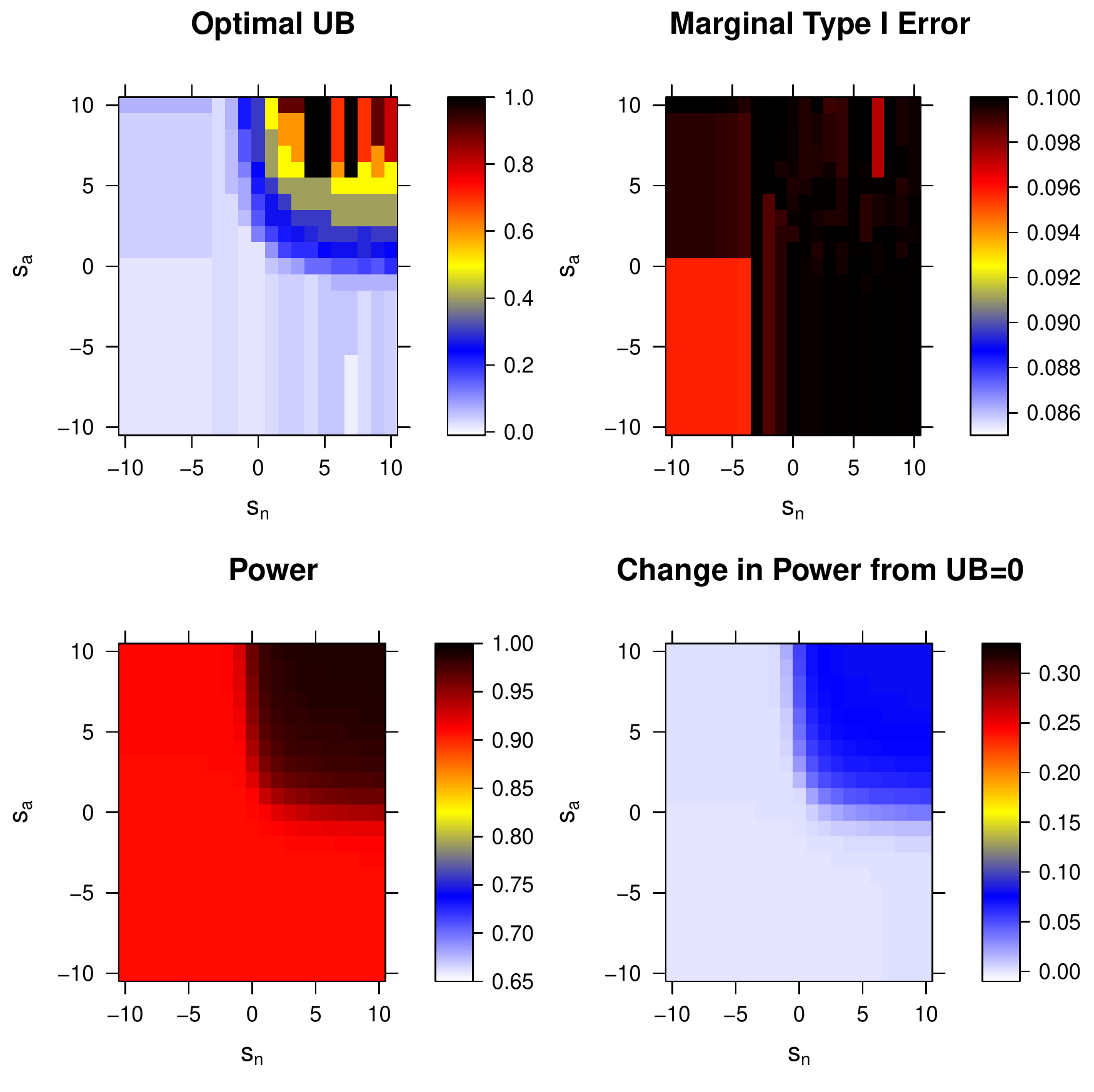}
\caption{Heat map of the optimal value of UB for 10 indication trial to maintain a weighted \textit{marginal} type I error less than or equal to 10\% for a given combination $s_{n}$ ($s$ for null indications) and $s_{a}$ ($s$ for alternative indications). Heat maps for weighted marginal type I error, weighted power, and change in power for the optimal UB compared to UB=0 (no borrowing).}
\label{fig:heatmap_2x2_marg_10bask}
\end{figure}

\begin{figure}[ht]
\includegraphics[scale=1]{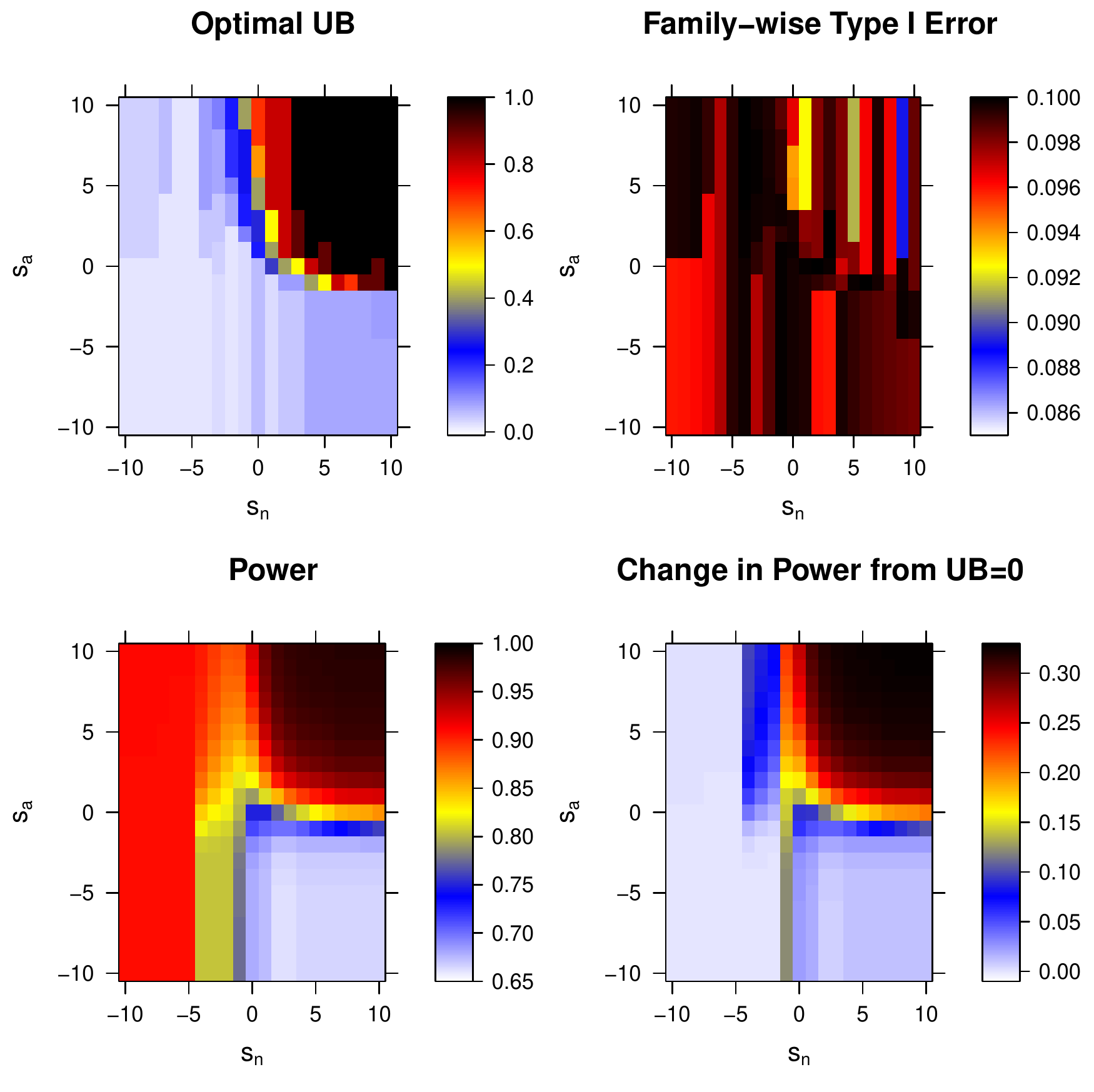}
\caption{Heat map of the optimal value of UB for 10 indication trial to maintain a weighted \textit{family-wise} type I error less than or equal to 10\% for a given combination $s_{n}$ ($s$ for null indications) and $s_{a}$ ($s$ for alternative indications). Heat maps for weighted family-wise type I error, weighted power, and change in power for the optimal UB compared to UB=0 (no borrowing).}
\label{fig:heatmap_2x2_fw_lte_10bask}
\end{figure}

Heat maps for the 10 indication trial simulations demonstrate similar results to the 5 indication trial considering marginal type I error rate control in Figure \ref{fig:heatmap_2x2_marg_10bask} or family-wise type I error rate control in Figure \ref{fig:heatmap_2x2_fw_lte_10bask}. The 10 indication results also demonstrate that at no combination of $s_{n}$ and $s_{a}$ is it optimal to not borrow (UB=0), but with less borrowing favored when $s_n$ or $s_a$ are negative.

\clearpage